# Mapping to a Reference Genome Structure

Benedict Paten[1], Adam Novak[1] and David Haussler[1,2]

1: Center for Biomolecular Sciences and Engineering, CBSE/ITI, UC Santa Cruz, 1156 High St, Santa Cruz, CA 95064, USA, 2: Howard Hughes Medical Institute, University of California, Santa Cruz, CA 95064, USA.

## Abstract

To support comparative genomics, population genetics, and medical genetics, we propose that a reference genome should come with a scheme for mapping each base in any DNA string to a position in that reference genome. We refer to a collection of one or more reference genomes and a scheme for mapping to their positions as a *reference structure*. Here we describe the desirable properties of reference structures and give examples. To account for natural genetic variation, we consider the more general case in which a reference genome is represented by a graph rather than a set of phased chromosomes; the latter is treated as a special case.

## Introduction

A genome assembly is typically represented as a set of strings over the nucleotide alphabet {A,C,G,T}, termed *contigs*, partitioned into a set of *scaffolds*, each of which is the concatenation of a sequence of contigs, interspersed with runs of wildcard characters (typically 'N') that represent uncertainty about the DNA sequence between the contigs.

A reference genome assembly is a genome assembly used to represent a species. The first draft of the human reference genome assembly (Lander et al., 2001) was monoploid, meaning that each chromosome was (essentially) represented by a single scaffold. Some polymorphic (variable) regions of the genome in the population were poorly represented by the chosen scaffold (Levy et al., 2007) (Wheeler et al., 2008) (Kidd et al., 2010). To better represent this variation, the current reference human genome assembly, GRCh38, while still primarily a chosen monoploid assembly, contains a substantial number of variant scaffolds, termed alternative haplotypes (Church et al., 2011). These scaffolds are mapped at both ends to the chosen monoploid assembly while deviating from it in the middle. GRCh38 can therefore be viewed as a graph, consisting of nodes representing reference DNA bases connected by edges that represent the main linear path of bases along the reference chromosome plus branching paths of polymorphism. In this paper we formalize this notion of a reference genome as a graph.

A central function of a reference genome is as a target for mapping DNA bases from other sequenced human genomes. For any position in an input string (typically a short "read" output directly from a sequencing experiment) a mapping to the reference genome is the identification of a position in the reference genome that is considered homologous. Such mappings provide information about the organization of the bases within the newly sequenced input genome, and, through analysis of the structure of the mapping, the variations that it contains.

There are four main problems with the current approach to representing the reference genome and mapping new input genomes to it. First, there is no standard way to map DNA bases from a newly sequenced input genome to positions in the reference genome, though a number of algorithmic implementations are very popular (Li & Durbin, 2009) (Ben Langmead, Trapnell, Pop, & Salzberg, 2009) (Zaharia et al., 2011). Second, mapping procedures have different ways of dealing with the case where, due to underlying repetition of larger subsequences in the reference, there are (nearly) equally good mappings to multiple disjoint locations within the reference (called the *multi-mapping problem*). Third, the GRC reference genome only incorporates a very limited amount of common segregating genome variation; the remaining variation is split among formats and data sources, e.g. the Single Nucleotide Polymorphism Database (dbSNP) (Sherry et al., 2001) and the 1000 Genomes Project (1000-Genomes-Project-Consortium, 2010). As a result, there is currently no single comprehensive resource for common human genome variation, and no single consistent naming or identification scheme for it. Finally, each time a new reference genome assembly changes, the coordinates of the reference genome change, and all data must be remapped. Mapping is often the most computationally expensive step in a genome analysis pipeline, and it can often takes weeks and consume significant resources to remap a large set of genomes.

Elaborating on the first problem, Figure 1 illustrates a typical case in which different mapping procedures may produce different results, here not in terms of overall read placement, which is also a very common difference, but rather regarding the placement of bases within a read. Here the base positions in a given input string are mapped (aligned) in equivalently optimal ways to the reference genome for different trades-offs between the number of mismatches and gaps (regions where one sequence does not map to the other). If a sequence variant is encoded using mapping to a reference, i.e. as an explicit difference from the reference genome, mapping ambiguity of this type frequently leads to a situation where the same input DNA sequence is treated as two different genetic variants by two different analysis pipelines, leading to higher level errors when interpreting associations between genotype and phenotype.

**(A)** (2 gap opens, 6 gaps, 0 mismatches)
```
Ref:    AAGCTA--CTAG----CT
Allele: AAGCTAGACTAGGAAGCT
```

**(C)** (1 gap open, 6 gaps, 2 mismatches)
```
Ref:    AAGCTA------CTAGCT
Allele: AAGCTAGACTAGGAAGCT
```

**(B)** (2 gap opens, 6 gaps, 0 mismatches)
```
Ref:    AAGCTA--CT----AGCT
Allele: AAGCTAGACTAGGAAGCT
```

**(D)** (1 gap open, 6 gaps, 2 mismatches)
```
Ref:    AAGCTACT------AGCT
Allele: AAGCTAGACTAGGAAGCT
```

**Figure 1. The problem of equivalently likely mapping.** Four different alignments of two sequences, "Ref" and "Allele", are shown, (A) and (B) both have 0 mismatches and 2 gaps, (C) and (D) both have 2 mismatches and 1 gap.

The reference representation and mapping approach taken by dbSNP goes some distance toward mitigating the four problems of the GRC reference genome approach, and is in many ways superior. It is much more comprehensive than the GRC reference in its catalog of variation, although it is restricted to short variants with no phasing information. Each genetic variant in dbSNP, including single nucleotide polymorphisms (SNPs), short insertions, deletions and other types of multibase allelic variants, is represented by a unique reference SNP ID (rsID), in conjunction with two flanking sequences of contextual DNA bases, one to the left (upstream) and the other to the right (downstream) of the variant position. The curators of dbSNP merge identical submissions so that the combination of an encoded variant nucleotide with upstream and downstream sequences is unique in the database, giving an unambiguous method for the identification of variants. The unique rsID record provides a shorthand numerical identifier for the variant that need never change.

Inspired by this approach, we propose an extension of the rsID scheme to define a reference structure that encompasses all bases of the human genome, both those that commonly vary and those that do not. A reference structure includes both a reference genome assembly and a scheme for mapping in which each position of the reference genome assembly is given a stable numerical identifier and a uniquely identifying set of context strings. To make this approach general, inclusive of large-scale as well as small-scale variation and fitting with the alternative haplotype model of GRCh38, we define this model using graphs, and show examples of how it can be used to more comprehensively and stably integrate common human variation.

## Results

### Sequence graphs

We model both reference and individual genome assemblies as *sequence graphs*, which allow for a more general class of assembly representation, and which incorporate phased contig/scaffold representations as a special case.

Base positions within a reference structure should have an aspect of permanence and universality, so that we can maintain canonical names and aliases for them indefinitely into the future. One possibility is to assign UUIDs to them for this purpose (http://en.wikipedia.org/wiki/Universally_unique_identifier), but other more compact identifiers can be used as well. A *base instance* is a pair ($b,P$) consisting of a labeling base $b$ in {A,C,T,G} and a *position P* (represented, e.g., as a UUID). All positions are globally unique, so given a position $P$, one may determine the base at that position, i.e. the unique $b$ such that ($b,P$) is a base instance. The nucleotides in all genomes are described as base instances.

In order to define a sequence graph, we need to specify how base instances are connected to form sequences. To do this generally, as DNA is double stranded, we must distinguish the forward and reverse complement orientations of base instances. Each base instance ($b,P$) has a *left side,* denoted $P_l$, and a *right side,* denoted $P_r$. An *adjacency* is an unordered pair of two sides;

an adjacency $\{P_s, Q_t\}$ asserts that the $s$ side of the base at position $P$ is connected to the $t$ side of the base at position $Q$.

A *sequence graph* $G = (V_G, E_G)$ is a bidirected graph (Medvedev & Brudno, 2009) in which each node in the set $V_G$ of nodes is a base instance and each edge in the set $E_G$ of edges is an adjacency connecting the sides of two base instances. The *forward label* of a node $(b,P)$ is the base $b$, and the *reverse label* is the reverse complement base $b^*$, where A* = T, T* = A, G* = C, and C* = G. Using its sides for orientation, for a base instance $(b, P)$ we write $b(P_l) = b$ and $b(P_r) = b^*$ to denote the base label oriented by the given side.

A *linear thread* is a special kind of path in a sequence graph composed of a sequence of oriented nodes and edges terminated by oriented nodes, such that each node other than the first and last node on the path is entered on one side and exited on the other. Nodes can be visited more than once in a thread. The *traversal* of a thread specifies a sequence of nucleotides, decoded by enumerating the labels of base instances in the order and orientation specified by the thread, such that if a base instance $(b, P)$ is oriented from $P_l$ to $P_r$ then $b(P_l)=b$ is incorporated into the traversal, and if oriented from $P_r$ to $P_l$ then $b(P_r)=b^*$ is incorporated into the traversal. A *circular thread* is a circular path of oriented nodes and edges in which each node is entered on one side and exited on the other. Its traversal is a circular sequence of nucleotides, e.g. a mitochondrial sequence.

A *contig (graph)* (we drop the word "graph" when it is clear from the context) is a sequence graph that consists of a single linear or circular thread with no node repetitions. A *phased sequence graph* is a sequence graph consisting of a set of disjoint contig subgraphs (Fig. 2). In Appendix A we discuss extension of (phased) sequence graphs and the mapping scheme presented below to represent complete linear chromosomes (sequences terminated at each end by special nodes called telomeres), and scaffolds (sequences of contig subgraphs interspersed with runs of Ns).

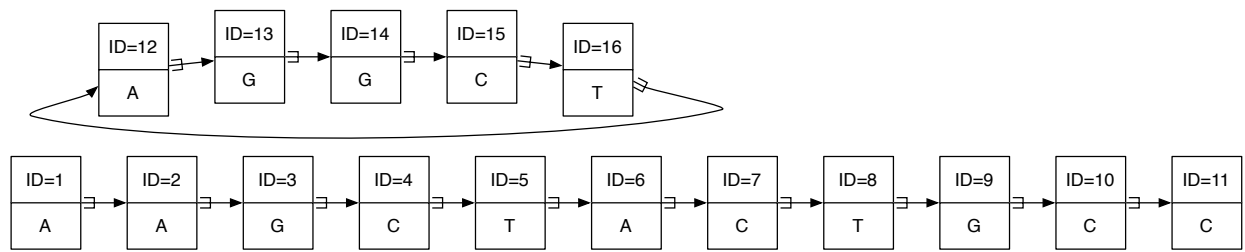

**Figure 2. A phased sequence graph with two contigs.** Each box is a node representing a base. The positions are denoted as "ID=n". In this and future figures, the adjacencies are depicted as lines terminated by arrowheads. The incident side for each adjacency endpoint is denoted by the shape of the arrowhead, with an outgoing arrowhead denoting the right side and incoming arrowhead denoting the left side, in common with earlier representations of bidirected graphs (Medvedev & Brudno, 2009). The short thread is circular, the long thread linear.

Any sequence graph that is not a phased sequence graph is called *unphased*. Unphased sequence graphs can be used to represent genomes in which there is some uncertainty in phasing or assembly. They can also be used to represent populations of genomes in which numerous variations are described, e.g. an extension of a reference genome assembly to include more than one variant of some regions (Fig. 3).

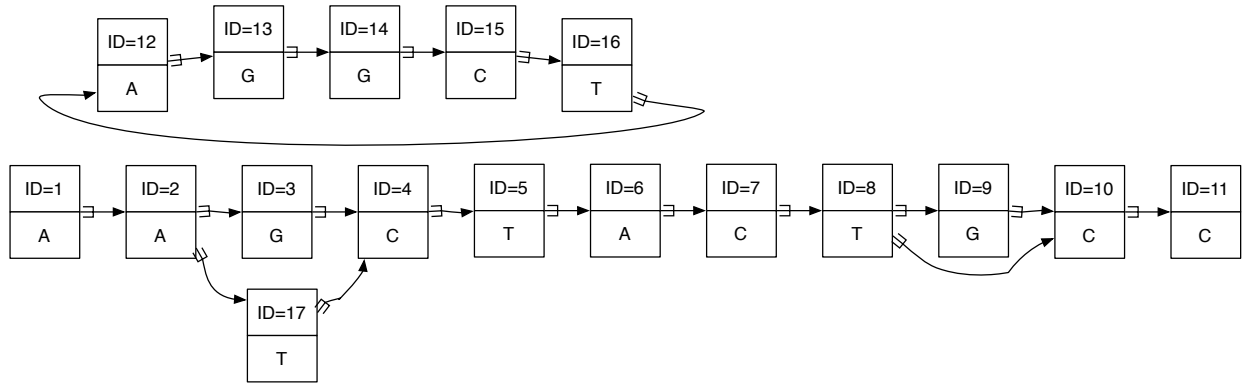

**Figure 3. An unphased sequence graph with two connected components.** This is a generalisation of Figure 1 in which allelic variations (a substitution and an indel) are shown.

## Comparing two sequence graphs

Let us assume that we are given an *input* sequence graph *G* and a *target* sequence graph *H*. The task is to map the input base instances in $V_G$ to corresponding target base instances in $V_H$. Often the input graph will represent the genome of a particular individual and the target graph will be used as a reference genome. To avoid wrongly categorizing genetic variations in the input graph, we leave a base instance in $V_G$ unmapped if it can plausibly map to more than one base instance in $V_H$. The mapping may therefore be partial; i.e. it is not assumed that all elements of $V_G$ will be mapped. Since the identifier *P* of the base instance (*b,P*) determines the base *b*, it suffices to map positions uniquely to positions, i.e. identifiers to identifiers. However, to account for the double-sided nature of DNA strands, we must allow a position *P* in *G* to map to a position in *H* in either the forward or reverse orientation. Formally, a *mapping* from a sequence graph *G* to a sequence graph *H* is a partial function *M* from the set of positions in *G* to the set of positions in *H* such that for every position *P* in *G*, either *M(P)* is undefined and we say that position *P* is *unmapped in H,* or there is a position *Q* in *H* such that *M(P) = Q* and either $b(P_l) = b(Q_l)$ (and we say that *P* is *forward mapped* to *Q in H*), or $b(P_l) = b(Q_r)$ (and we say that *P* is *reverse mapped* to *Q in H*). In either of the latter two cases we say that position *P maps to the position Q in* H (or more generally that the position *P* is *mapped in H*).

Example: Mapping to a Phased Sequence Graph Using Left-right Exact String Match

Let *H* be a phased sequence graph to be treated as a target (e.g. reference) genome. For each side $Q_s$ in *H* the *unique context* of $Q_s$, denoted $U(Q_s)$, is the shortest suffix *u* of the traversal of a thread in *H* ending at *Q* entered from its *s* side, such that *u* occurs exactly once as a traversal of a thread in *H* (ending at any side of any node). The suffix *u* includes the base of Q, and does not always include other bases. If no such unique suffix *u* exists for $Q_s$, we say that the side $Q_s$ is *unmappable* (Fig. 4).

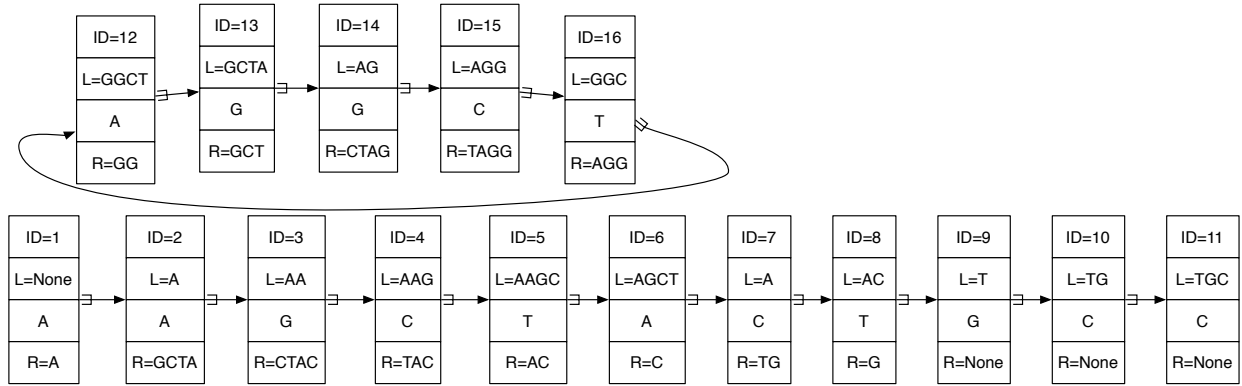

**Figure 4. Left and right unique contexts.** Shown here are the unique contexts for every side in the sequence graph from Figure 1. For a node *(b, P)* denoted with *L=t* and *R=u* the left unique context is the string *tb* and the right unique context is the string *[bu]\**; in this and subsequent figures the right unique context string is shown as its reverse complement so that both unique contexts of a position can be read from left-to-right. A side labeled "None" is unmappable.

For a side $P_s$ in a phased sequence graph G, $P_s$ is *matched* to a side $Q_t$ in a phased target sequence graph H if a suffix of the traversal of the thread ending at and including P entered from its *s* side is the same as $U(Q_t)$. The essential property of the unique contexts of H is that, as a family, they don't share suffixes, which means any side in any *G* can be matched to at most one side in *H*.

Given the above matching of sides, the *left-right exact match mapping* $M_e$ of positions is defined as follows (Fig. 5); a position P in G is:
- *unmapped* in H (and $M_e(P)$ is undefined) if neither $P_l$ or $P_r$ are matched anywhere,
- *left-mapped* at $M_e(P) = Q$ if $P_l$ is matched to a side of Q, but $P_r$ is not matched anywhere,
- *right-mapped* at $M_e(P) = Q$ if $P_r$ is matched to a side of Q, but $P_l$ is not matched anywhere,
- *fully-mapped* at $M_e(P) = Q$ if $P_l$ is matched to one side of Q and $P_r$ is matched to the other side of Q,
- else P is *inconsistently-mapped* in H (and $M_e(P)$ is undefined).

Combining cases, we say that position P is *mapped in H* if it is left-mapped, right-mapped or fully-mapped, and *not mapped in H* if is it either unmapped or inconsistently-mapped. This defines the left-right exact match mapping $M_e$ from G to H.

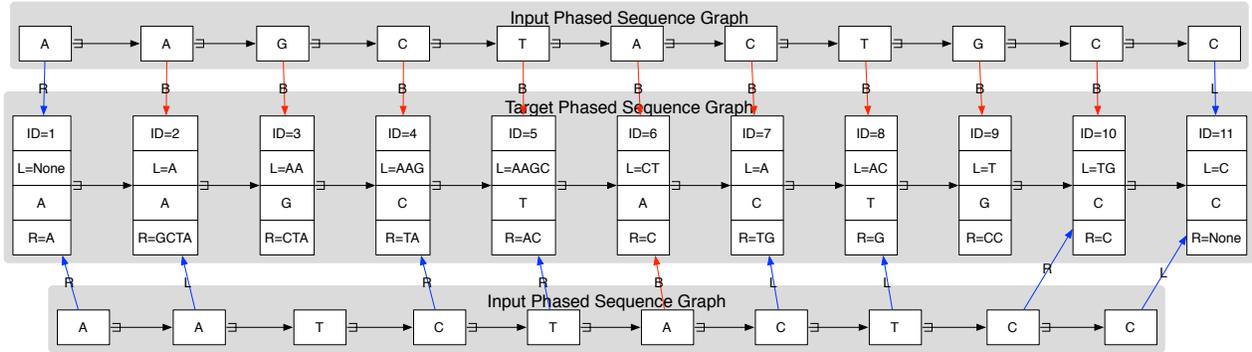

**Figure 5. Mapping phased input sequence graphs to a phased target sequence graph for the left-right exact match mapping case.** The input sequence graphs are shown without IDs, the blue/red arrows indicate the mapping. Arrows either labeled with an L, indicating left mapping, an R, indicating right mapping, or a B, indicating full mapping. The input sequence graph above the target sequence graph is isomorphic to the target sequence graph, and the mapping reflects this. The input sequence graph below the target sequence graph contains variations with respect to the target.

## General context-driven mapping schemes

A *mapping scheme for a sequence graph H* is a definition, rule or procedure that defines a mapping *M* from positions in *G* to positions in *H* for all possible sequence graphs *G*. The pair *(H, M)*, consisting of a sequence graph *H* and a mapping scheme for *H* is called a *reference structure*. For example, if *H* is a phased sequence graph and we define the mapping *M* from *G* to *H* to be the mapping $M_e$ for each phased sequence graph *G*, and leave the mapping of every position in a unphased input graph *G* undefined, then *M* is a mapping scheme for *H* and (*H, M*) is a reference structure.

In general, we would like nontrivial mapping for unphased sequence graphs as well, for increased sensitivity we may want to allow some mismatches or indels at individual bases when comparing strings for similarity, we may want to use two-part contexts that include both bases to the left and the right of the position to which we are mapping, and for greater stability we may want to use contexts that are longer than the minimal required context. In these cases we need to allow for more flexibility in defining the contexts of a position in a target sequence graph, while retaining the essential properties of left-right unique exact matching.

A (*two-part*) *context* for a position *Q* in a sequence graph *H* is a triple $c = (L, b_l(Q), R)$ of DNA sequences with *L* referred to as the *left part* and *R* as the *right part* such that its *string* $s(c) = L\ b_l(Q)\ R$ is the traversal of a thread in *H* passing through *Q* from left to right, where in this traversal $b_l(Q)$ is preceded by *L* and followed by *R*. The context $c' = (L', b_l(Q'), R')$ of the position *Q'* is *forward more general than* the context $c = (L, b_l(Q), R)$ of the position *Q* if $b_l(Q) = b_l(Q')$, *L'* is a suffix of *L*, and *R'* is a prefix of *R*. The context *c'* is *reverse more general than c* if $b_l(Q) = b_r(Q')$, *L'* is a suffix of *R\**, and *R'* is a prefix of *L\**. We say *c' is more general than c* if it is either forward or reverse more general than *c*. A collection of sets of contexts is *nonredundant* if when collapsed into a single multiset of contexts, no context is more general than any other. A *context assignment,* **C***,* for a sequence graph *H* is defined by specifying for each position *Q* in *H* a nonempty *context set* $U_C(Q)$ of contexts such that the resulting family $\{U_C(Q): Q \text{ is in } H\}$ is nonredundant.

Let $H$ be a target sequence graph and $C$ be a context assignment for $H$. For any position $P$ in any sequence graph $G$ and any position $Q$ in $H$, we say that $P$ *forward exact matches* $Q$ if there is a context of $P$ in $G$ that is in $U_C(Q)$. Similarly, the position $P$ *reverse exact matches* $Q$ if there is a context $(L, b_l(P), R)$ of $P$ in $G$ whose reverse complement $(R^*, b_r(P), L^*)$ is in $U_C(Q)$. We say that $P$ *exact matches* $Q$ if $P$ forward or reverse exact matches $Q$. In Appendix C we define more general notions of matching that include forms of inexact matching. For simplicity, in the main text we will henceforth use the terms "matching" and "exact matching" interchangeably, always meaning exact matching.

If position $P$ in input sequence graph $G$ matches to one and only one position $Q$ in target sequence graph $H$ we define $M_C(P) = Q$ and say that *P maps to Q* in $H$, else $M_C(P)$ is undefined in $H$ and we say that P is *not mapped* in $H$. It is clear that if $P$ maps to $Q$ then in order to match $b_l(P)$ to $b_l(Q)$ or its reverse complement, all matches of $P$ to $Q$ must be in the same direction. Therefore, we can further specify a mapping to be either a forward mapping or a reverse mapping. In the case where $P$ is not mapped, we can also further distinguish the subcase where $P$ does not match any position in $H$ ($P$ is *unmapped*) or $P$ matches more than one position in $H$ ($P$ is *inconsistently mapped*).

Every context assignment $C$ to the positions in $H$ defines the mapping scheme $M_C$ and a reference structure $(H, M_C)$. A mapping scheme (reference structure) defined in this way is called *context-driven*.

It is clear from the way that matching is defined that if a context $c$ in the context set for position $Q$ in $H$ were allowed to be more general than another context $c'$ in the context set for position $Q'$ in $H$, then any input position $P$ in any input graph that matches $Q'$ using $c'$ would also match $Q$ using $c$. Thus, if $Q$ equaled $Q'$ then $c'$ would be superfluous, and if $Q$ did not equal $Q'$, then any match to $c'$ would create an inconsistent mapping, so $c'$ would be useless. This is the reason that the context sets in a context-driven mapping scheme are required to be nonredundant. Since each context set is also required to be nonempty, nonredundancy also implies that for any position $Q$ in $H$ we can always construct an input graph $G$ with a position $P$ that matches $Q$ by giving P a context that does not contain a context for any other position in $H$. In particular, if $G = H$ then each position maps to itself.

Because of these strict conditions, not every sequence graph has a context-driven mapping scheme. For example, it is easy to see that a phased sequence graph has a context assignment if and only if it has neither one chromosome that is contained in another, nor a circular chromosome of the form *W, W, ..., W* with more than one repetition of the DNA word *W* (see the Appendix D for further discussion). In practice, this is not a serious limitation. We call a sequence graph that has a context-driven mapping scheme a *mappable* sequence graph.

A context assignment can either be defined explicitly as a collection of sets of pairs of strings or implicitly as a computational procedure that determines whether any string *w* is *s(c)* for some two-part context in the assignment (or has a suffix, prefix of subword that is *s(c)*).

Example: General Left-right Exact Match Mapping Scheme

We say a context *c = (L,$b_l$(Q),R)* for a position *Q* in *H* is *unique* if it is not also a context for any other position in *H* (forward or reverse) and that it is *minimally unique* if it cannot be shortened by removing a base at the beginning of *L* or at the end of *R* without ceasing to be unique. We say that *c* is a *right context* if *L* is the empty string, and *c* is a *left context* if *R* is the empty string. A sequence graph in which every position has either a left unique context or a right unique context is called *left-right mappable*. These are a subclass of mappable sequence graphs, e.g. the sequence graph formed from the 3-contig phased genome {ACT, CTG, TGA} is mappable but not left-right mappable because the "T" in "CTG" requires context on both sides for uniqueness. For any left-right mappable phased target graph *H*, the mapping defined in the previous section can be seen to be a context-driven mapping scheme as follows: we assign at most 2 contexts for each position *Q* in *H*, one left context *c = (L,$b_l$(Q),R)* with an empty right part *R* and a left part *L* that is just long enough to make its string *s(c) = L $b_l$(Q)* minimally unique, and the other a right context *c' = (L',$b_l$(Q),R')* with an empty left part *L'* and a right part *R'* that makes its string *s(c) = $b_l$(Q) R'* similarly minimally unique.

It is straightforward to generalize the above left-right exact match mapping scheme to arbitrary input sequence graphs and arbitrary left-right mappable target sequence graphs that are not necessarily phased. For each position *Q* in *H* we just include in $U_C$(Q) a left minimally unique context *c = (L,$b_l$(Q),R)* with an empty right part *R* for every possible thread in *H* entering on the left and ending at *Q*, and we do a similar thing with empty left part for threads beginning at *Q* and extending to the right. We call this the *(general) left-right exact match mapping scheme*. This is a very practical and flexible mapping scheme. Figures 6 and 7 show examples of general left-right exact match mapping for, respectively, phased and unphased input sequence graphs. In Appendix C we discuss a related general left-right inexact context-driven mapping scheme, and in Appendix B a natural scheme with left, right and additional "in between" contexts.

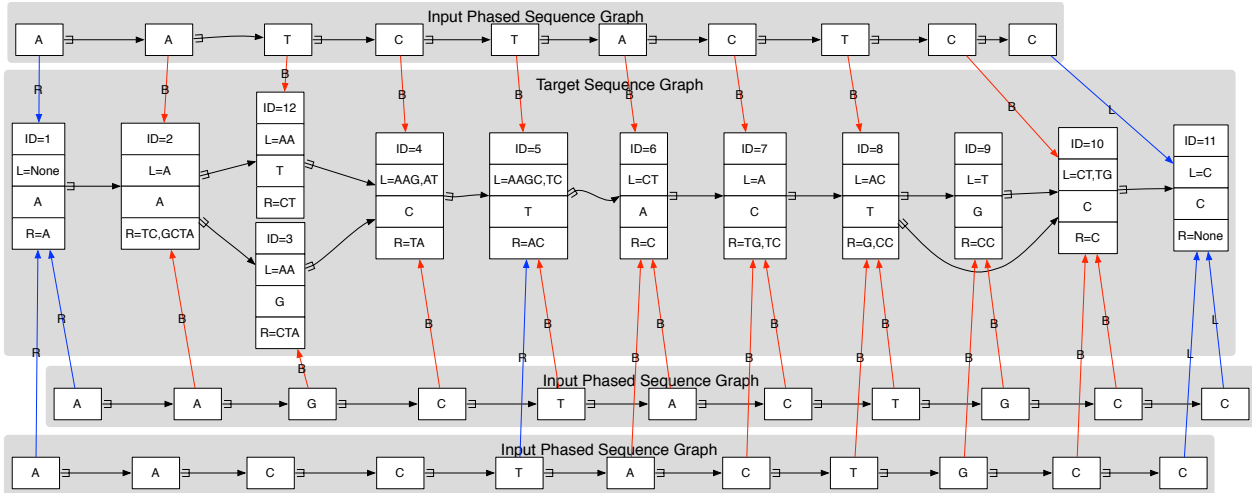

**Figure 6. Mapping phased input sequence graphs to a target unphased sequence graph for the left-right exact match mapping case**. The two haplotypes used to build the target sequence graph are shown as the input sequence graphs immediately above and below the target sequence graph. Mapping is denoted as in Fig. 4. L and R are now left and right minimally unique contexts for all threads (excluding the base itself). A novel haplotype that differs from the haplotypes used to build the target sequence graph is shown at the bottom as an input sequence graph mapped to the target sequence graph.

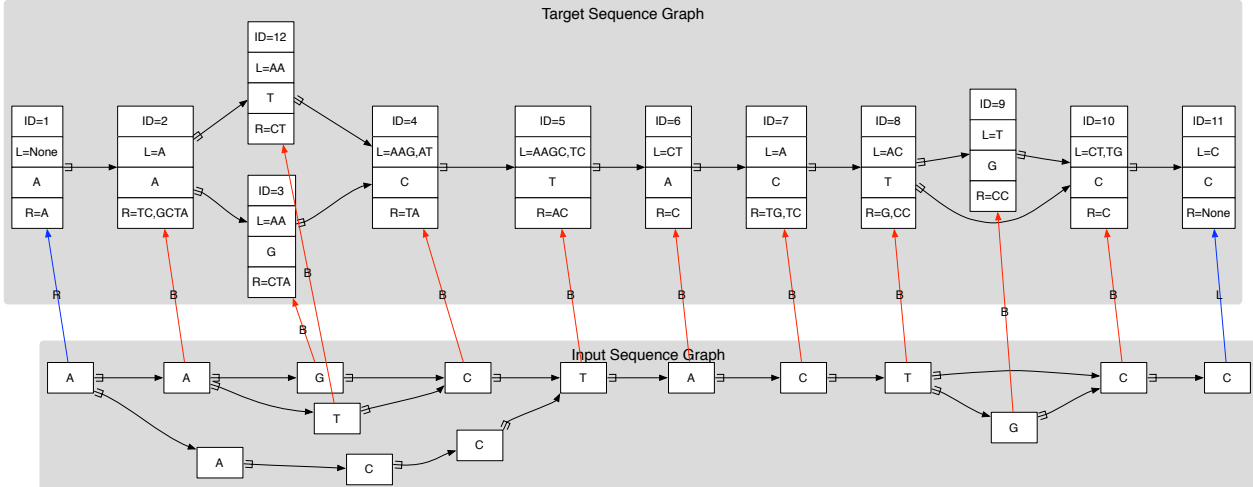

**Figure 7. Mapping an unphased input sequence graph to a target unphased sequence graph for the left-right exact match mapping case**. The three haplotypes shown as distinct input sequence graphs in Fig. 5 are merged together into a single input sequence graph (see further discussion of merging in a subsequent section) and mapped to the target sequence graph (above).

## Example: Central Exact Match Mapping Scheme and de Bruijn Mapping

The opposite extreme from the imbalance of left-right exact match mapping scheme is a scheme called the *central exact match mapping scheme* in which the contexts are minimally unique and we attempt to make the left and right parts of each context have the same length. A *central minimally unique context* for a position $Q$ in $H$ is a minimally unique context $c = (L, b_l(Q), R)$ for $Q$ such that the absolute value of the difference between the length of $L$ and the length of $R$ is as small as possible among all minimally unique contexts of $Q$. The context assignment for the central exact match mapping scheme includes just the (at most two) central minimally unique contexts in the context set for each position. For example, let $H$ = {TAGACTACGCT} be construed as a phased sequence graph. Then for the position $Q$ in the middle, labeled T, the left minimally unique context string is ACT, the right minimally unique context string is TAC, and there is one central minimally unique context string CTA. For the position $Q"$ at the rightmost end, also labeled T, the left minimally unique context string is GCT, this is also the only central minimally unique context, and there is no right minimally unique context. Finally, for the rightmost position labeled G, call it $Q'$, the left minimally unique context string is CG, the right is GC, and these are both also central minimally unique contexts.

A variant of the central exact match mapping scheme is the *balanced central exact match mapping scheme* in which it is required that the left and right parts of each context have the same length and the combined context be unique (but not required to be minimally unique). The best example of a balanced central exact match mapping scheme is a *k de Bruijn mapping scheme* for an odd length $k = 2p+1$. This mapping scheme is defined on a sequence graph $B_k = B_k(W)$ called a *k* de Bruijn graph (Pevzner, Tang, & Waterman, 2001) (Zerbino & Birney, 2008) that has one node for each double-stranded segment of DNA of length $k$ from a given contig library $W$. The sequence graph $B_k$ has an adjacency edge between two nodes $a$ and $b$ if $b$ is a 1 base shift of $a$, i.e. if $b$ is obtained by removing a base-pair from one end of $a$ and adding another base pair to the other end of $a$, and the k+1 base string obtained by overlapping k-1 bases of $a$ and $b$ is in $W$. Arbitrarily choosing a top strand for every node, if we write the top strand of the double-stranded DNA segment represented by the node with position $Q$ as $LbR$, where $b = b_l(Q)$, and $L$ and $R$ are the flanking DNA sequences of length $p$, then the $k$ de Bruijn context assignment $B$ is $U_B(Q) = \{(L,b,R)\}$. This defines the de Bruijn mapping scheme $M$ for $B_k$, and $(B_k,M)$ is called a *k de Bruijn reference structure*. For any position $P$ in a phased input graph $G$ that is in the center of a *k*-mer that appears in $W$ (on one strand or the other), the $p$ flanking bases to the left and right determine the position $Q$ in $B_k$ to which $P$ maps. If $P$ is not in the center of a *k*-mer that appears in $W$ then $P$ does not map to $B_k$. An example de Bruijn reference structure is shown in Figure 8.

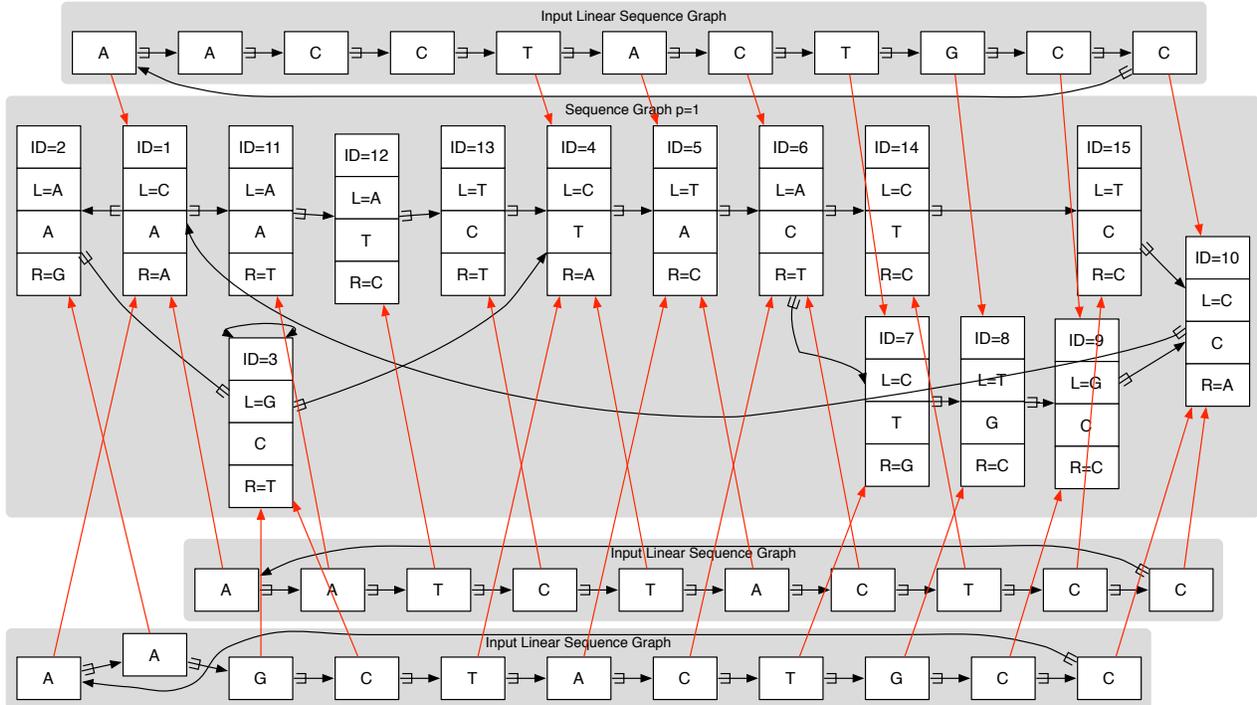

**Figure 8. Mapping phased input sequence graphs to a target unphased sequence graph for the *k de Bruijn mapping scheme*, where here k=3.** Other than the difference in mapping scheme, the figure is laid out as in Figure 6, with the difference that here the input sequences are circular, ensuring that all positions are mappable with unique contexts of length k=3.

## Context-driven reference structures and reference hierarchies

A pair ($G,M_C$) consisting of a sequence graph $G$ and a context-driven mapping scheme $M_C$ for $G$ defined by context assignment **C** is called a *context-driven reference structure*. Given context-driven reference structures ($G,M_C$) and ($H,M_D$), position $P$ in $G$, and position $Q$ in $H$, we say that $P <= Q$ if for every context $u$ in $U_C(P)$ there is exactly one context $v$ in $U_D(Q)$ that is more general than $u$, and no context more general than $u$ in $U_D(Q')$ for any other position $Q'$ in $H$. We say that ($H,M_D$) *is more general than* ($G,M_C$), written ($G,M_C$) <= ($H,M_D$) or, for short, $G <= H$, if for each position $P$ in $G$ there exists a position $Q$ in $H$ with $P <= Q$. As this position $Q$ must be unique by definition, when $G <= H$ we write $Q = f_H(P)$ and refer to $f$ as the generalization function between $G$ and $H$. The relation <= defines a partial order on the families of context sets of context-driven reference structures, i.e. it is reflexive, transitive, and antisymmetric. In particular, if $G <= H <= I$, then $G <= I$ and $f_I(f_H(P)) = f_I(P)$. We call a set of reference structures that are partially ordered by <= a *(context-driven) reference hierarchy.*

Given context-driven reference structures ($H,M_C$) <= ($H',M_{C'}$), position $Q$ in $H$, and position $Q' = f_{H'}(Q)$ in $H'$, then any context of any position $P$ of any input graph $G$ matching a context of $Q$ in $U_C(Q)$ also matches a unique corresponding (possibly more general, i.e. shorter) context of $Q'$ in $U_{C'}(Q')$. Thus, if $P$ matches $Q$ in $H$, then P matches $Q' = f_{H'}(Q)$ in $H'$
It follows further from the transitivity of <= itself that for any chain $(H_1,M_1), …, (H_m,M_m)$ of reference structures from a reference hierarchy such that $(H_1,M_1) <= … <= (H_m,M_m)$, for any position *P* in any input sequence graph *G,* and any *i*, if *P* matches position *Q* in $H_i$, then for any *j* >= *i,* P matches $f_j(Q)$ in $H_j$, where $H_j$ as a subscript is abbreviated as *j*.  In other words, anything that matches to a position in an earlier, less general reference structure in the chain will match to the corresponding more general position in any subsequent more general structure in the chain. We call this *transitivity of matching*.

Matching does not imply mapping. If *P* maps to *Q* in $H_i$  (i.e. *P* matches just the position *Q* in $H_i$), then either *P* maps to $f_j(Q)$ in $H_j$ (i.e. *P* matches just the position $f_j(Q)$ in $H_j$) or *P* does not map at all in $H_j$ because it matches more than one position in $H_j$. To determine the mapping of *P* to higher levels in the chain, we can follow pre-computed links from *P* defined by the generalization function *f* between positions at consecutive layers to find the first matching, but we must also check for new matchings in this higher level in positions other than the one found by following the pre-computed link.

However, if we are given the information that *P* matches to position *Q* in some reference structure in a reference hierarchy, then the only possible positions in less general reference structures from that hierarchy that *P* could match are those positions in these structures that are less general than *Q*. There may be only a few such "candidate less general matching positions", and they can be rapidly recovered by following the same pre-computed links for the generalization function *f* in the opposite direction, say in a depth-first search. At each candidate matching position, a test could be performed to see if *P* does indeed match. If not, this branch of the depth-first search can be terminated, as transitivity of matching implies that when *P* fails to match to any position *Q* in a reference structure, then it will also fail to match to any position in any less general reference structure in the hierarchy. Therefore, by determining the matches to the most general reference structure(s) in a reference hierarchy, we can employ the generalization function *f* to find all matches in all other structures in the hierarchy, and by checking if there is a unique match in a given reference structure, determine all mappings as well.

**Context-driven merging schemes**

The most natural way to obtain a more general reference structure is to merge nodes. Two base instances (*Q, b*), (*Q', b'*) that are nodes in a sequence graph *G* can be *forward merged* if *b* = *b'*, by replacing them with a base instance (*Q'', b*), redirecting every adjacency edge that contains $Q_l$ or $Q'_l$ to attach instead to $Q''_l$, and doing the similar thing on the right side. The *reverse merge* is similar except the left and right sides of *Q'* are reversed, and *b\** must equal *b'*. The result of a series of node merges is invariant to ordering, so this notion of a merge is readily extended to 3- and higher-way merges. We can also allow the positions being merged to come from different sequence graphs by first forming a single graph consisting of the disjoint union of the sequence graphs, then performing node merges.

A (*context-driven) merging scheme* is a function (method) that takes a collection of context-driven reference structures **H** = {($H_1,M_1$), …, ($H_m,M_m$)} and produces a *merged reference* structure *(H, M)*, such that *(H$_i$, M$_i$)* <= *(H, M)* for all *1 <= i <= m*. Many merging schemes are possible.

## Example: overlap merging scheme

A simple, parameterless merging scheme, which we call the *overlap merging scheme,* is defined as follows. We say that positions *P* in *H$_i$* and *Q* in *H$_j$* are *overlapping* if there is a context in **U**$_i$*(P)* that is more general than a context in **U**$_j$*(Q)* or vice versa, where here we denote a context assignment by its index. The *overlap graph* of **H** is the graph whose nodes are the positions in graphs of **H**, and that has an edge between each pair of overlapping positions. The *merged sequence graph H"* of **H** is obtained by merging all the nodes in each connected component of the overlap graph. The direction for merging in this process is never ambiguous (i.e. it is always determined to be either a forward or reverse merge), and the two sides of a single position are never merged because they cannot be overlapping, as the forward and reverse complement base of a base instance are always distinct.

The *minimization C'* of a context assignment **C** in a context-driven reference structure *(H, M$_C$)* is defined as follows. For each base instance *(Q, b)* in *H* and every family of contexts *F = {(L,b,RR$_1$)*, …,*(L,b,RR$_n$)}* in **U**$_C$*(Q)* such that *c = (L,b,R)* is not more general than any other context for *Q* or for any other position in *H*, and for which *R* is the shortest right part prefix that when paired with the left part prefix *L* has this property, we replace *F* in **U**$_C$*(Q)* with the single context *c*, and we do similar substitutions on the left sides. This is followed by a *clean-up merge* in which any positions with overlapping context sets (after minimization) are merged as described above. The resulting reference structure *(H', M$_{C'}$)* is called *the minimization of (H, M$_C$)*.

Finally, the *overlap merge* of a collection of context-driven reference structures **H** = {*(H$_1$,M$_1$)*, …, *(H$_m$,M$_m$)*} is the reference structure *(H, M)* obtained by first computing the merged sequence graph *H"* and setting the context assignment **C"** such that the context set at a merged position is the union of all context sets for positions merged together to form that merged position, and then setting *(H, M)* to be the minimization of *(H", M$_{C''}$)*. It can be verified that *(H, M)* is a context-driven mapping scheme and *(H$_i$, M$_i$)* <= *(H, M)* for all *1 <= i <= m*.

## Example: *p*-overlap merging schemes

For some purposes it is desirable that strings in either the left or right part of any context be no longer than a specified (positive integer) length *p*; this is particularly useful when mapping reads whose length is less than or equal to *p+1* when using left-right matching, or *2p+1=k* when using central matching.

We say that positions *P* in *H$_i$* and *Q* in *H$_j$* are *left p-overlapping* if they are overlapping or there is a context *(L,b$_l$(P),R)* in **U**$_i$*(P)* such that *L* shares a suffix of length p with the left part of a context in $U_j(Q)$ and $b_l(P) = b_l(Q)$, or $L$ shares a suffix of length p with the reverse complement of the right part of a context in $U_j(Q)$ and $b_l(P) = b_r(Q)$. Similarly, *P* and *Q* are *right p-overlapping* if they are overlapping or there is a context $(L, b_l(P), R)$ in $U_i(P)$ such that *R* shares a prefix of length p with the right part of a context in $U_j(Q)$ and $b_l(P) = b_l(Q)$, or *R* shares a prefix of length p with the reverse complement of the left part of a context in $U_j(Q)$ and $b_l(P) = b_r(Q)$. We say *P* and *Q* are *p-overlapping* if they are either left or right p-overlapping, and that they are *symmetrically p-overlapping* if they are both left p-overlapping and right p-overlapping.

For a set of context-driven reference structures **H**, each with a context-driven-mapping scheme, the *(symmetric) p-overlap merge* produces a merged reference structure *(H, M)* by first merging all *(symmetrically) p*-overlapping positions and then proceeding as in a typical overlap merge. A reference hierarchy can be produced by starting with simple context-driven schemes for the least general reference graphs, such as the central exact matching scheme, and applying (symmetric) *p*-overlap merges for various values of *p* to produce more general reference graphs (Figure 10). Alternatively, a mixture of merging schemes can be applied successively to achieve a reference hierarchy. For example, after applying the overlap merge scheme to a set of input graphs, a *p*-overlap merge can be applied at higher levels to ensure that the left and right strings of all contexts are less than or equal in length to *p* (Figure 10).

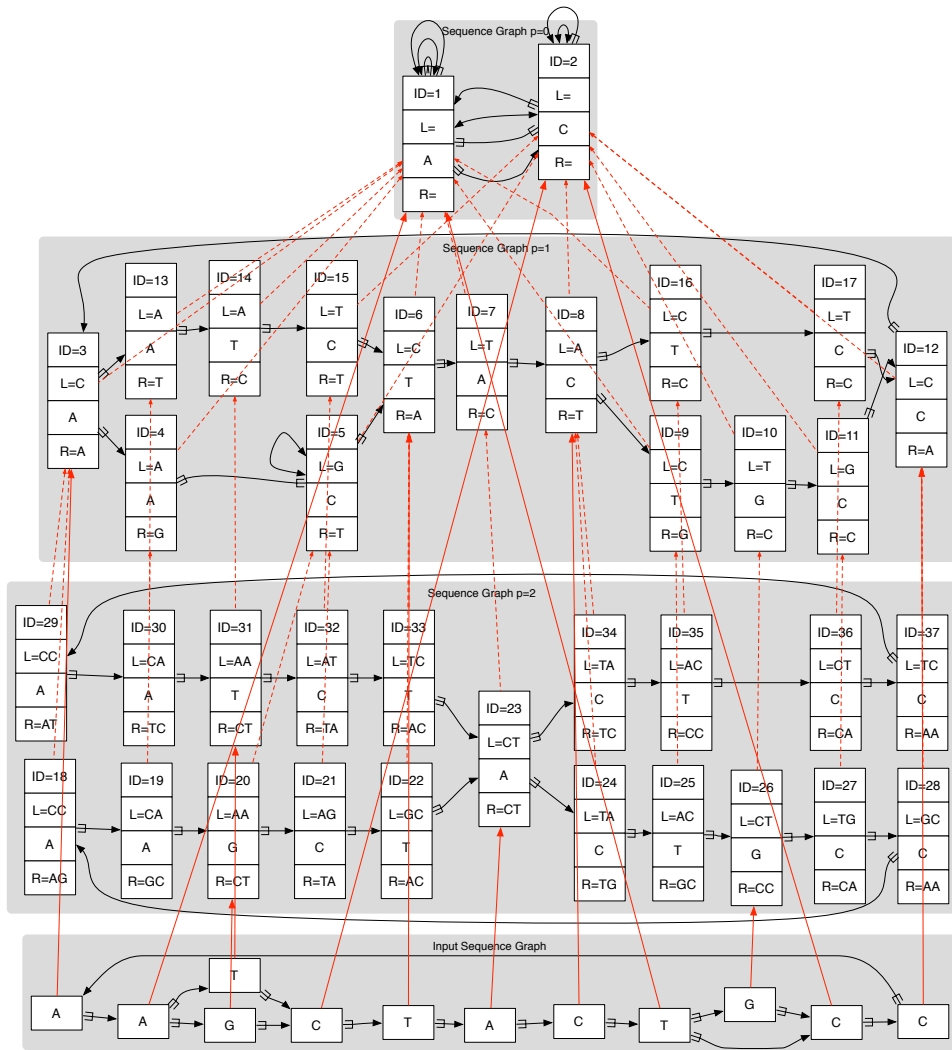

**Figure 9. A reference hierarchy constructed using the *symmetric p-overlap merging scheme, with an input sequence graph mapped to it using the central exact match mapping scheme.*** Starting from a phased sequence graph with contexts defined by a *central exact match mapping scheme,* a symmetric *p*-overlap merging scheme with the indicated decreasing value of *p* was used to form each layer. All the graphs in the reference hierarchy are de Bruijn graphs. Dotted red lines show mapping between positions in the hierarchy, while solid red lines show mapping of the input sequence graph into the hierarchy.

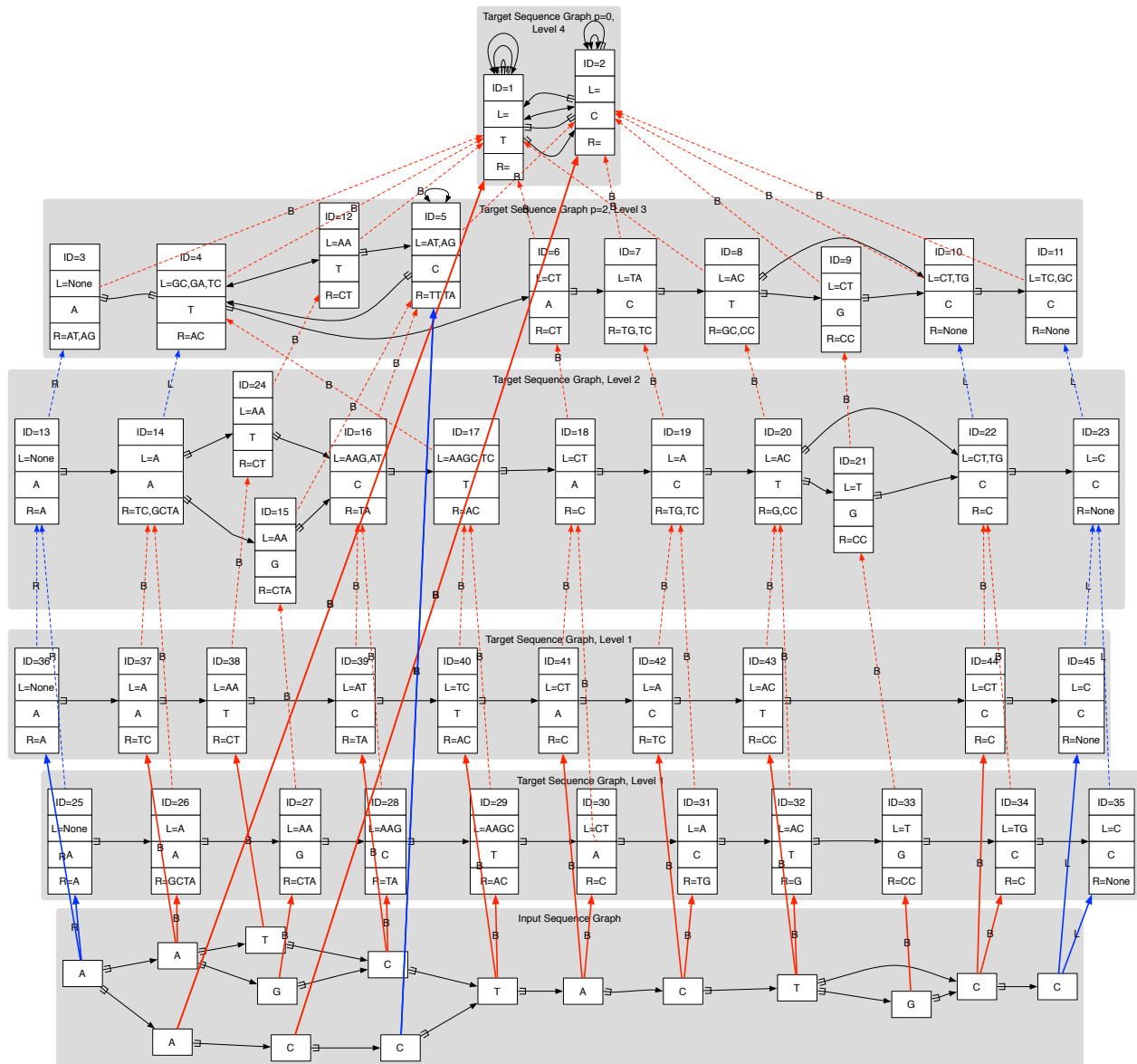

**Figure 10. Mapping an input sequence graph to a mixed reference hierarchy using exact left-right matching.** Level 1 is composed of two contig graphs, level 2 merges the level 1 graphs together using the overlap merging scheme. To build level 3, a p-overlap merge scheme for p=2 is applied. The final level is built by using the p-overlap merge scheme for p=0.

# Discussion

We've introduced a scheme to define reference structures with positions that have persistent identifiers, and with the ability to both represent a wide spectrum of human genetic variation and provide an integral method for mapping. This scheme avoids the problem of ill-defined alignment to the reference. In addition, we've shown how the multi-mapping problem can be dealt with by creating a hierarchy of reference structures. We've defined reference structures using sequence graphs, and described context-driven mapping schemes that employ simple exact string matching. This is concordant with the indexing schemes for (directed acyclic) graphs (Sirén,

Välimäki, Longi, & Mäkinen, n.d.), which build upon the Burrows Wheeler Transform (M Burrows, 1994), and future implementation of the reference structures we introduce here is likely to use such schemes.

A reference hierarchy built from human genomes organized by populations and subpopulations would be a good way to create a rigorous representation of segregating human genetic variation for both population and medical genomics. Such a reference hierarchy could start from many individual human genomes at the bottom level, each in a separate context-driven reference structure. These could then be grouped into larger and larger subpopulations at higher levels, each such subpopulation being represented by a merged sequence graph containing all the variation present in the subpopulation's bottom level genomes. The depth-first search process for rapidly recovering mappings can be used in a hierarchy of this type to find haplotypes in subpopulation-specific reference genomes that match a position in any given human input genome.

A de Bruijn or similarly merged reference, perhaps at a higher level in the hierarchy, can similarly be used to map positions in the repetitive areas of a reference genome from shorter contexts. If a phased reference genome contains a piece of DNA larger than a read size $p$ that is repeated multiple times as identical paralogs, then these will be merged into a single subgraph the $p$-overlap merged graph. While it would require a very large context to map uniquely to a position in this repetitive region in an unmerged reference, in a $p$-overlap merged reference a position would be uniquely mapped with a typical read-sized context. Once we map to the unique position in the merged reference using the short context, we can use the depth-first search procedure to efficiently recover all the separate paralogs in the unmerged genomes at lower levels in the hierarchy that were merged to form this single position in the merged reference.

Handling unmapped positions

Some positions in any new input genome may represent novel elements of that genome, such as virally inserted or highly mutated stretches of DNA that cannot be reasonably mapped to positions in even the most general reference genome in the reference hierarchy. The positions in an input genome *G* that fail to map to a most general reference genome *H* in a reference hierarchy are called *unmapped positions*, and the rest of the positions are called *mapped positions*. Let *U(G)* be the sequence graph that is obtained from the input graph *G* by removing all the mapped positions and their adjacencies. We call *U(G)* the *unmapped subgraph of G* and we call the connected components of *U(G)* the *unmapped components* of *G*. Each unmapped component *U* from *U(G)* is adjacent to a set of mapped positions in *G* that we denote *F(U)*, where *F* stands for "flanking nodes". The flanking nodes in *F(U)* collectively map to a set of nodes in *H*. These positions are called the *neighborhood of U*. While we can't map the positions in the unmapped component *U* to positions in *H*, we can at least "approximately map" them to the neighborhood of *U* in *H*.

An important special case occurs when the input sequence graph *G* is phased. In this case there are at most 2 nodes in *F(U)* for any unmapped component *U* of *G*. Thus, the neighborhood of *U* is at most 2 nodes in *H*, and the unmapped component of *G* defines a single novel path in *H* that can be envisioned by adding nodes to *H* corresponding to those in *U*. This novel path either spans two nodes in *H* if there are two nodes in the neighborhood, hangs off the one node in *H* if there is only one node in the neighborhood, or exists entirely on its own as a separate component if there are no nodes in the neighborhood. Hence, using this strategy, we can think of the unmapped component *U* as a piece of novel or highly mutated DNA that is inserted at (or provides an alternate at) a particular place in the reference genome *H*. Since there is a canonical way of identifying this novel variant represented by *U* in terms of the DNA sequence of *U* and the at most 2 nodes in *H* that constitute the neighborhood of *U*, if an equivalent variant *U'* is observed again in a second genome will have the exact same form, and can be recognized as being homologous to *U*. At this point the newly discovered recurrent variant might be unambiguously named as a new type of human genetic variant if so desired, which might later be added to the graph representing the human reference structure, simplifying future handling of it.

Representing rearrangements of an input genome relative to a reference hierarchy

The mapped positions in an input sequence graph will not always be in the same relative order and orientation to each other as they are in a reference genome from the reference hierarchy. If we select (or create) a phased reference genome *H* based on information in the reference hierarchy, and compare it to a phased input genome *G*, then the difference in relative order and orientation between *G* and *H* among the mapped positions of *G* can be represented in a kind of breakpoint graph, often used in comparative genomics (Pevzner, 2000). We get a standard breakpoint graph when there is at most one position in the input genome *G* mapping to any single position in the reference genome *H*. Here we simply elide any intervening unmapped positions in *G* or *H*, so the resulting mapping is 1-1 and onto. The breakpoint graph represents the net changes due to the rearrangements between the input genome *G* and the reference genome *H* that have occurred with regard to the mapped positions. These net changes are naturally decomposed into cycles or chains of alternating colored edges in the breakpoint graph, with one color representing adjacency in the input genome *G* and the other representing adjacency in the reference genome *H*, again ignoring the elided positions where there was no mapping. Simple changes such as inversions, translocations, etc., are easily identified in a breakpoint graph. There is a rich theory of such graphs, which can be exploited to identify and analyze structural variation in a mapped phased input genome.

Comparing two input genomes mapped to the same reference genome

Assume that two phased input sequence graphs *G* and *G'* are mapped to the same phased reference sequence graph *H*. If *G* and *G'* are large and most of the positions in each are mapped, before doing anything else we might like to quickly check if *G* and *G'* are isomorphic, i.e. if they represent the same genomes as double-stranded DNA. Assume that for any pair *(P,Q)* of positions in *H* we can quickly determine if *G* and *G'* both map continuously to the reference interval in *H* between *P* and *Q*, and if so, determine the number of segments in *G* and *G'* that map continuously to this interval. If this number of segments that map to the designated interval in *H* is not the same for both *G* and *G'*, then *G* and *G'* are not isomorphic. Otherwise we are finished dealing with this part of *G* and *G'* and our isomorphism problem is now decomposed into separate isomorphism problems for the corresponding portions of *G* and *G'* outside of the already confirmed mapped regions. By proceeding with such a divide-and-conquer strategy, eventually getting down to direct comparison of short strings, we can quickly determine whether or not there is isomorphism between *G* and *G'*.

Updating a reference structure to a new version

One of the biggest annoyances (and resource drains) in genomics is all the remapping that must be done whenever the official human reference genome assembly is updated to a new assembly. In the approach we advocate here, in which we maintain a hierarchy of reference structures, this is less of a problem. Each identifier that identifies a specific position in a sequence graph in the hierarchy is a permanent identifier that will never need to be changed. So long as we allow the reference genomes in the hierarchy to be sequence graphs representing human reference variation, follow a discipline of only adding additional nodes and edges when we update them, and only add new reference genomes as needed at the bottom of the hierarchy, never subtracting anything, we can minimize the amount of remapping that needs to be done when a new version of the official reference hierarchy is released. This is discussed in Appendix E.

# Summary

Reference hierarchies of sequence graphs and accompanying context-driven mapping schemes combine the strengths of the GRC reference genome with those of the dbSNP variation catalog to provide a single unified approach to human genomic reference variation. As the uses of genomics in science and medicine rapidly expand in the coming decade, it is vital that we take time to re-examine our methodology for defining human genomic reference sequences and variants, so that we can have a system that is both comprehensive and efficiently extensible, while remaining computationally scalable. Reference hierarchies offer an attractive approach.

# Acknowledgements

We thank Heng Li, Gad Getz, Ewan Birney and Richard Durbin and other members of the Global Alliance for Genomics and Health Data Working Group for helpful conversations and providing examples that motivated the described approach. We in particular thank Heng Li for providing the alignment example in Figure 1.

# Appendices

## Appendix A: Representing linear chromosomes and scaffold ambiguity

In finished or near-finished genomes, such as the current human reference genome, complete linear chromosomes are represented as near-complete strings. To represent a complete linear chromosome as a thread each side of a linear thread not incident with an adjacency can be connected by adjacencies to a distinct *telomere node* that has an adjacency only on one side and has a pseudo-base represented by the telomere label "$". It is unnecessary to distinguish the forward and reverse complements of telomere labels, therefore $ = $*. A telomere node is not counted as a base instance, but it does have a position allowing it to be uniquely identified. A contig in which every base instance has a single adjacency on one side is a *chromosome (graph)*.

Scaffolds are concatenations of contigs with regions of sequence uncertainty. They are typically represented by inserting sequences of wildcard ('N' character) symbols between adjacent contigs. One easy way to represent scaffolds in the context of a reference genome graph is to simply insert a fixed number of positions labeled 'N', and interpret this label as representing any of the 4 possible bases. If it is necessary to represent regions of sequence uncertainty within a sequence graph in a more flexible manner, we propose labeling adjacencies. Let the width of an adjacency *a*, denoted *w(a)*, be a triple (*min(a), med(a), max(a)*) of integers with $0 <= min(a) <= med(a) <= max(a)$ that if unspecified all default to 0, representing a direct connection between the sides in *a*. In general, if $a = \{ P_s, Q_t \}$, the integers *min(a), med(a),* and *max(a)* are the lower bound, median and upper bound, respectively of the number of unspecified intervening bases between the $P_s$, and $Q_t$. Formally, a thread is a *contig (graph)* if all its adjacencies have (upper bound) width zero, else it is a *scaffold (graph)*.

In defining the traversal of a thread containing adjacencies with non-zero width we can include a gap by including a sequence of 'N' characters, such that for each adjacency *a* in a thread we insert a label of *med(a)* 'N' characters, enumerating the labels of the base instances and adjacencies in the order of thread (with special traversal rules for bases as before) to form the traversal string.

Typically 'N's in DNA sequences are treated as ambiguity characters, considered identical when matching to any member of the nucleotide alphabet. However, for defining contexts it is undesirable to treat them as ambiguity characters, because this obviously leads to all context strings having length longer than the length of the longest contiguous run of N characters in any traversal of the graph. To avoid this, for the purposes of defining context assignments, we would propose treating 'N' characters as not matching any base A, C, G or T, so that 'N's are therefore excluded from context strings, and are unmapped when present in an input genome that is mapped to a reference genome. Subsequent procedures can then be defined on a per-application basis to define how unmapped 'N's are handled downstream of the primary mapping functions.

## Appendix B: The total exact match mapping scheme

The left-right exact match mapping scheme makes use of only the minimally unique contexts that are either shifted maximally toward the left part of the context or shifted maximally toward the right part. It is also possible to include additional contexts that have nonempty left and right parts, such as the contexts in the central exact match mapping scheme. Adding additional contexts gives more opportunities for matching, allowing some positions in some input sequence graphs to map that were previously not mapped. However, it can also make a position inconsistently mapped that was previously (consistently) mapped, so there is a tradeoff here.

The most comprehensive context assignment composed of minimally unique contexts includes in the context set $U_C(Q)$ every minimally unique context $c = (L,R)$ for $Q$. This is called the *total exact matching context assignment*. For each position Q in H, the context set $U_T(Q)$ of the total exact matching assignment *T* contains the context set for $U_C(Q)$ the left-right exact matching scheme *C*. The additional contexts in $U_T(Q)$ can be found by an "inch worm crawl" between an extreme left context (*L,e*) and an extreme right context (*e,R*) from $U_C(Q)$, where *e* is the empty string. This crawl begins on the extreme left with *L' = L* and *R'= e*. Then in a front extension and back contraction step, the shortest prefix *R"* of *R* is found such that there exists a proper suffix *L"* of *L* such that the context (*L",R"*) is minimally unique. This context (*L",R"*) is then added to $U_T(Q)$ and the crawl is repeated starting at (*L",R"*) until we reach a point where *L" = e* (and necessarily *R" = R*).

## Appendix C: Mapping schemes that allow inexact matching

In the main text, for simplicity, while defining a general context-driven scheme, we considered only examples involving exact sequence matches, however, context schemes are easily defined that feature substitutions and indels; here we define one such scheme.

Let *H* be a sequence graph and let *j, k* be nonnegative integers. We say that two strings are *similar* if one can be transformed into the other by at most *j* substitutions and *k* single-base insertions or deletions on the remaining prefixes. Similarity is therefore a bounded form of edit distance. A (*two-part*) *similar context* for a position *Q* in a sequence graph *H* is a triple (*L,b_l(Q),R*) of DNA sequences such that there exists a context (*L',b_l(Q),R'*) for Q in *H* and *L* is similar to *L'* and *R* is similar to *R'*. For any exact match context driven mapping scheme we can define its analogous *j,k* similarity based inexact matching mapping scheme, simply replacing contexts with similar contexts when defining the (similar) context sets. The tradeoff is that the length of minimally unique similar contexts will grow substantially with increasing *j* and *k* and the number of similar contexts will grow exponentially with *j* and *k*. Positions may also become unmappable due to multi-mapping, even though the graph contains no trivial automorphisms, as discussed in the next appendix.

## Appendix D: Automorphisms and subgraph automorphisms

An *isomorphism* between a sequence graph *G* and a sequence graph *H* is a 1-1 onto mapping between the sides of nodes of *G* and the sides of nodes of *H* such that adjacencies are preserved and each node in *G* either maps to a same-labeled node in *H* preserving left and right sides, or to a reverse-complement labeled node in *H* swapping left and right sides. If *G = H* this mapping is called an *automorphism*, and is *nontrivial* if not every node forward maps to itself. If *H* is phased with no circular chromosomes, then it has no nontrivial automorphisms if and only if all its chromosomes are distinct as double-stranded DNA. If we allow circular chromosomes, then we must additionally require that no circular chromosome has the form *WW ... W^* for two or more repetitions of a nonempty word *W*.

A *subgraph isomorphism* from *G* to *H* is an isomorphism from *G* to a subgraph of *H*. If *G* in fact is a subgraph of *H*, then a subgraph isomorphism is called a *subgraph automorphism*, and is nontrivial if not every node in *G* forward maps to itself. If *H* is phased, then it has no nontrivial subgraph automorphisms for any of its connected components if and only if no chromosome is contained in any other chromosome as double-stranded DNA, and no circular chromosome has the form *WW ... W^* for two or more repetitions of a nonempty word *W*. This is the case if and only if *H* is mappable.

## Appendix E: Updating a reference structure to a new version

Consider a hierarchy of context-driven reference genomes with left-right exact match mapping at the bottom level of the hierarchy, and each level derived from the one below by a defined (*p*)-overlap merging strategy. Let *H* be an old sequence graph in the bottom level, and *H'* be a new replacement sequence graph for *H* obtained by adding nodes and edges to *H* in such a way that H' has a left-right exact match mapping scheme. Extend the generalization function *f* by defining the function *f* from positions in *H* to corresponding positions in *H'* be the 1-1 identity function on the positions of *H*, leaving the other new positions in *H'* out of range.

Let *G* be any input sequence graph. By the definition of the left-right exact match mapping scheme and the fact that *H'* is obtained from *H* by adding nodes and edges, for any position *P* in *G* that maps to position *Q* in the old sequence graph *H*, if *P* maps at all to the new graph *H'*, the only place that *P* can map is the same position *f(Q)*. We just need to check if *P* actually does map to *f(Q)* in *H'*, e.g. by looking at the context sets for *f(Q),* the only thing that may have changed. Often the unique context sets for *f(Q)* in *H'* will be the same as they were for *Q* in *H*. We call *Q* an *unchanged position* in *H* (and in H') in this case. Any position in any input sequence graph that maps to an unchanged position *Q* in *H* and also maps to some position in *H'* must map to the position *f(Q)* in *H'*. We can keep a table of the unchanged positions and know that we never need to update the mappings to them. For a small update to the reference sequence graph, the vast majority of the positions will likely be unchanged.

If *P* in *G* maps to *Q* in *H* but does not map to *f(Q)* in *H'* then this is may be because a suffix *u* in a unique context set of *Q* in *H*, one that was formerly used to map *P* to *Q*, is no longer in the unique context set of *Q = f(Q)* in *H'* because it is no longer unique in the expanded reference sequence graph *H'*, and is (because *H'* retains an exact match mapping scheme) in fact replaced with one or more extensions of the suffix *u* that no longer match any suffix of the traversal of a thread ending at *P*. This can be systematically detected. If there remains in a unique context set of *f(Q)* a suffix *u* with a match from *P*, the other reason that *P* in *G* might map to *Q* in *H* but not map to *f(Q)* in *H'* is because the addition of nodes and edges to *H* in forming *H'* creates an extra match of a different context of *P* to a different node in *H'* . This can also be systematically detected.

On the other hand, if *P* does not map to *Q* in *H* but does map to *f(Q)* in *H'*, then this is because there is a new suffix in the context set of *f(Q)* derived from a thread that uses the added positions in *H'* that was not in any of the relevant context sets of positions in *H* (and in particular, not in the relevant context set of *Q*). In either case, we can say that *Q* is an *extended position* whose new context sets are obtained by lengthening the strings in the previous contexts sets, or adding new strings to the previous context sets. As above, we can keep a table of the extended positions (every position in *H* is either unchanged or extended) and use the details of the particular extension of the context sets of an extended position *Q* to decide if any position *P* that previously mapped to *Q* no longer maps there, or if any position *P* that previously did not map anywhere in *H* now maps there.

The last things we need to pay attention to are the context sets for the new positions in *H'*. As above, we can use these to check if any position *P* that previously did not map anywhere in *H* now maps to one of these new positions in *H'*.

These are the tests that must be done to update the mapping of any input sequence graph *G* to every changed genome in the bottom of the reference hierarchy **H**. The merged sequence graphs at the higher levels of **H** are automatically recomputed from the altered sequence graph set at the bottom level by applying the merging scheme, thereby forming the updated reference hierarchy **H'**. The mapping from an input sequence graph *G* to the higher-level sequence graphs in **H'** is updated as follows. By the transitivity of matchings, if a position *P* in *G* matches to any position *Q* anywhere in **H',** and in particular at the bottom level, then *P* matches to all more general positions *Q'* such that *Q <= Q'*, and these can be found by tracing the generalization links via the function *f* for **H'.** As discussed above, we also have to check for additional matches to other positions in these higher levels. Conversely, if *P* fails to match to a position *Q* anywhere in **H'**, then *P* fails to match to *Q'* for all positions *Q' <= Q,* again by the transitivity of matchings. These relations can be used to narrow down the positions in the hierarchy to which *P* matches, and from there the mapping may be determined.

Assume identifiers are UUIDs. When additions are made to the bottom level of the reference hierarchy, new UUIDs are created for the new base positions added there. The positions in the merged graphs at the higher levels are treated in a special way. First note that in any hierarchical reference structure, since each position at a level higher than 1 is obtained from merging one or more nodes at the previous layer, the generalization function *f* defines a forest of UUIDs, with increasing generality as one travels toward the root of any tree. One can think a UUID of an internal node as representing the set of UUIDs for the leaf nodes below it (in Level 1). As we update the reference hierarchy as described above, this property is preserved, and moreover, every new UUID we create at any level higher than 1 after the update now represents some union of the sets of UUIDs previously represented at this level, plus possible some additional new UUIDs added to Layer 1: namely, the UUIDs for the nodes that were merged to form this new UUID. No set of UUIDs every represented previously by a higher level UUID is ever split. Therefore the history of UUIDs at the higher levels also forms a tree structure in time, each new UUID pointing back to the previous UUIDs and new Layer 1 UUIDs from positions that were merged to create it, effectively, it points back to its former aliases.

This defines a natural scheme for keeping track of mapping provenance. It makes any new nomenclature for human variation defined from the new, expanded reference hierarchy backwards compatible with the nomenclature defined from previous versions.